\documentclass[fleqn,12pt]{SelfArx} 

\usepackage{lipsum} 

\usepackage{listings}

\usepackage{fancyvrb}

\usepackage{pdfpages}

\usepackage{graphicx}

\usepackage{amsmath}

\definecolor{color1}{RGB}{0,0,90} 
\definecolor{color2}{RGB}{0,20,20} 

\usepackage{hyperref} 
\hypersetup{hidelinks,colorlinks,breaklinks=true,urlcolor=color2,citecolor=color1,linkcolor=color1,bookmarksopen=false,pdftitle={Title},pdfauthor={Author}}

\JournalInfo{Independent Research } 
\Archive{Spring, 2014} 

\PaperTitle{ Medical Volume Rendering Techniques } 

\Authors{Wenhui Zhang\textsuperscript{1}*, James K. Hahn\textsuperscript{1}} 
\affiliation{\textsuperscript{1}\textit{Department of Computer Science, The George Washington University, Washington D.C., United States of America}} 

\affiliation{*\textbf{Corresponding author}: wenhui@gwmail.gwu.edu} 

\newcommand{\keywordname}{Keywords} 

\Abstract{Medical visualization is the use of computers to create 3D images from medical imaging data sets, almost all surgery and cancer treatment in the developed world relies on it.Volume visualization techniques includes iso-surface visualization, mesh visualization and point cloud visualization techniques, these techniques have revolutionized medicine. Much of modern medicine relies on the 3D imaging that is possible with magnetic resonance imaging (MRI) scanners, functional magnetic resonance imaging (fMRI)scanners, positron emission tomography (PET) scanners, ultrasound imaging (US) scanners, X-Ray scanners, bio-marker microscopy imaging scanners and computed tomography (CT) scanners, which make 3D images out of 2D slices. The primary goal of this report is the application-oriented optimization of existing volume rendering
methods providing interactive frame-rates. Techniques
are presented for traditional alpha-blending rendering, surface-shaded display, maximum
intensity projection (MIP), and fast previewing with fully interactive parameter control. Different preprocessing strategies are proposed for interactive iso-surface rendering and fast previewing, such as the well-known marching cube algorithm.
 }

\begin{document}

\flushbottom 

\maketitle 

\tableofcontents 

\thispagestyle{empty} 


\section*{Introduction} 

\addcontentsline{toc}{section}{Introduction} 

Medical imaging currently plays a crucial role throughout the entire clinical applications from medical scientific research to diagnostics and treatment planning. However, medical imaging procedures are often computationally demanding due to the large 3D medical datasets to process in practical clinical applications.

In this report, we gave a compact 30-year overview of the history of medical visualization research.  Based on the histroy we catergorized the whole survey report into direct volumne rendering and indirect volume rendering techniques. Detailed view about volume rendering in general is also descibed.  Software only acceleration methods are presented as well.

Research challenges of the coming decade are identified and discussed in the last section.


\section{Thirty Year Overview of Medical Volume Visualization}

Since the off spring of magnetic resonance imaging (MRI) and computed tomography (CT) scanners around the early 70s, and the consequent tons of medical volume data, medical visualization has undergone significant development and is now a primary branch of visualization.

Medical visualization includes the use of various techniques to either invasively or noninvasively image the structure/anatomy and function/pharmacology of human being bodies. Structural medical visualization explains physical properties of the recording system, while functional medical visualization reveals physiological constraints of the recording system. Regular clinical methods for creating images and visualization of human being structures include Computed Tomography (CT), Structural Magnetic Resonance Imaging (MRI), Endoscopy, Ultrasound Image (US) while regular clinical methods for creating images of human being function includes Electroencephalography(EEG), Magnetoencephalography(MEG), Positron Emission Tomography (PET) and Functional Magnetic Resonance Imaging (fMRI).

Medical Visulization finds its applications in diagnostic purposes and treatment purposes, therefore medical visulization has an important role in the improvement of public health in all population groups. For example virtual colonoscopy, in treatment, for example surgical planning and guidance, and in medical research, for example visualization of diffusion tensor imaging data  \cite{Preim} .  As importance of medical visualization is revealed in both the acute and postacute rehabilitation settings for patients, like brain visualization with ruptured aneurysm brain tumor, lesions and other common 560 brain diseases like Alzheimer's, Parkinson's, Altruism, and Anxiety Disorder, 3D medical visualization interactive tools needed to be built up to meet the satisfaction of  current therapeutic and diagnostic clinical standards.

Furthermore, medical visualization is justified also to follow the course of a disease already diagnosed and treated. During the past decades, there was rapid developemnt in medical image acquisition technology, which make it now possible to acquire much more complex data than human beings can acheive ever before. Take High Angular Resolution Diffusion Imaging (HARDI) for example, 40 or more diffusion weighted volumes are acquired, thus made it possible to calculate and visualize water diffusion and, indirectly, structural neural connections in the brain \cite{Tuch} . In fMRI based full brain connectivity, time based correlation of neural activity is indirectly measured between all pairs of voxels in the brain, thus giving insight into the functional neural network  \cite{Greicius} . Moreover, advanced medical visulization technics enforces us on answering more and more complex answers.

%

In 1978, Sunguroff and Greenberg published their work on the visualization of 3D surfaces from CT data for diagnosis, as well as a visual radiotherapy planning system, also based on CT data \cite{Sunguroff}.

 In 1983, Vannier et al. published their results on developing a system for the computer based preoperative planning of craniofacial surgery  \cite{Vannier} .  The system was based on the extraction and visualization of 3D hard and soft tissue surfaces from CT data. Through the integration of an industrial CAD (computer aided design) application, it was also possible to perform detailed 3D measurements on the extracted surfaces.

In 1986, Hohne and Bernstein published a paper on using the gray-level gradient to perform shading of surfaces rendered from 3D CT data \cite{Hohne}.

 In 1987, Lorensen and Cline published the now famous Marching Cubes isosurface extraction algorithm  \cite{Lorensen}, which enabled the fast and practical extraction of 3D isosurfaces from real world medical data.

In 1988,  Levoy publised a paper on volume raycasting \cite{Levoy_raycasting}.

In 1988,  the first multimodal volume rendering paper was  published by Hohne \cite{Hohne1988}, in which registration and combined visualization of CT and MRI was introduced and nicely presented.

One of the first real applications of medical visualization is known as therapy planning, which remains important to this day. In 1993, Altobelli published their work on using CT data to visualize the possible outcome of complicated craniofacial surgery \cite{Altobelli1993}.

Basser published a paper in 1994, introducing Diffusion Tensor Imaging ( DTI ),  an MRI based acquisition modality, which yields $3*3$ symmetric diffusion tensors as its native measurement quantity \cite{Basser1994}.

In 1995, Hong introduced virtual colonoscopy (VC) \cite{Hong1995}, after which, medical visualization is serving as an more and more important medical application, namely screening for colon cancer.

Time varying medical volume data visualization was brought on the table in 1996 by Behrens,  to support the examination of dynamic contrast enhanced MRI mammography data with the display of parameter maps, the selection of regions of interest (ROIs), the calculation of time intensity curves (TICs), and the quantitative analysis of these curves \cite{Behrens1996}.

In 1998, Basser and his colleagues published one paper on extracting data from fibertract trajectories from DTI data, they were known as the first to extract and visualize fibertract trajectories from DTI data of the brain at that time\cite{Basser1998}. After two years, the visualization community includes tensor lines for tractography  \cite{Weinstein1999} and direct volume rendering of DTI data \cite{Kindlmann1999}  \cite{Kindlmann2000}  spring out with tons of innovative methods.

In 2000 Ebert introdeced the term illustration in this work \cite{Ebert2000}. Illustrative visualization is primarily motivated by the attempt to create renditions that consider perceptual capabilities of humans. Boundary enhancement based on gradient approximation \cite{Csebfalvi2001} and curvature-based transfer functions \cite{Kindlmann2003}  are landmarks in illustrative medical visualization. Tietjen et al. applied silhouettes and other feature lines for various scenarios in liver surgery planning \cite{Tietjen2005}. Besides silhouettes technique,  hatching has great potential to reveal details of shapes  \cite{Interrante1995} .

In 2001, Tory presented methods for visualizing multi time pointed MRI data of a multiple sclerosis patient, where the goal was to study the evolution of brain white matter lesions over time, which sets the milestone for multi subjects medical visualization \cite{Tory2001}.

For a long time, it was not possible to apply illustrative visualization techniques in practice due to performance constraints. And in 2003, GPU raycasting was introduced by Kruger \cite{Kruger2003}, with advances in graphics hardware and algorithms, it is now feasible from a computational standpoint.

The upcoming technology of DTI initiated a whole body of medical visualization research dedicated to the question of how best to visually represent and interact with diffusion tensor data in particular and multifield medical data in general. In 2007, Blaas presented a paper on a visual analysis inspired solution to this problem based on linked physical and feature space views \cite{Blaas2007}.

Medical visualization has also started to work on the problem to work with multi subjects data. These datasets include measurements and imaging for more than one subject at a time. The aim of this paper is to be able to extract and analysis certain patterns that affect subgroups of the whole collection.  LifeLines2, an early information visualization system to visualize and compare multiple patient histories or electronic medical records \cite{Wang2008}. Work has been done on the interactive visualization of the multi subject and mixed modality datasets acquired by medical cohort studies \cite{Steenwijk2010}. In these studies, mixed modality data, including imaging, genetics, blood measurements, is acquired from a group of subjects in order to be anaylized for diagnosing or predicting the clinical outcome of that group. It was demonstrated by Stenwijk to create a highly interactive coupled view visualization interface, integrating both information and scientific visualization techniques, with which patterns, and also hypotheses, could be extracted from the whole data collection.

 Area of medical visualization is very complex and, depending on a context, requires supplementary activities of medical doctors, medical physicists, biomedical engineers as well as technicians.

\section{Volume-rendering Methods}
Volume rendering is a technique for visualizing sampled functions of 3D  data by computing 2D projections of a colored semitransparent volume.It involves the following steps: the forming of an RGB-Alpha volume from the data, reconstruction of a continuous function from this discrete data set, and projecting it onto the 2D viewing plane (the output based on screen space) from the desired point of view. 

The raw datasets we got for medical purpose include, cloud point data, data by slides(nii file in  neuroscience field, e.g. MRI), surface data, to show these dataset in volumetric way needs some special technics and transfer functions to transfer them into RGB-Alpha dataset modality. 

An RGB-Alpha volume is a 3D four-vector data set, where the first three components are the familiar R, G, and B color components and the last component, Alpha, represents opacity. An opacity value of 0 means totally transparent and a value of 1 means totally opaque. Behind the RGB-Alpha volume an opaque background is placed. The mapping of the data to opacity values acts as a classification of the data one is interested in. Isosurfaces can be shown by mapping the corresponding data values to almost opaque values and the rest to transparent values. The appearance of surfaces can be improved by using shading techniques to form the RGB mapping. However, opacity can be used to see the interior of the data volume too. These interiors appear as clouds with varying density and color. A big advantage of volume rendering is that this interior information is not thrown away, so that it enables one to look at the 3D data set as a whole. Disadvantages are the difficult interpretation of the cloudy interiors and the long time, compared to surface rendering, needed to perform volume rendering.

 There are four main paradigms in which volume rendering is performed in nowadays: raycasting  \cite{Levoy1988} \cite{Tuy1984} , splatting \cite{Westover1990} , shear warp \cite{Lacroute1994} , cell projection \cite{Max1990} \cite{Shirley1990} , texture mapping hardware assisted \cite{Cabral1994} \cite{Engel2001} \cite{Scheltinga1995}, and via custom hardware \cite{Meissner1998} \cite{Pfister1999}.

In this report, we are only interested in software based volume rendering technics, and volume rendering techniques are clustered into two catergorites, indirect volume rendering and direct volume rendering. Indirect volume rendering, where in a preprocessing step the volume is converted to an intermediate representation which can be handled by the graphics engine. In contrast, the direct methods process the volume without generating any intermediate representation assigning optical properties directly to the voxels.

\subsection{Indirect volume rendering}

Indirect volume rendering technique extracts polygonal surface from volume data and represents an isosurface, it is also known as as 3D contours. The most popular algorithm for indirect volume rendering is marching cube algorithm \cite{ Lorensen}. 

Indirect methods aim at the visualization of isosurfaces defined by a certain density threshold. The primary goal is to create a triangular mesh which fits to the isoregions inside the volume. This can be done using the traditional image processing techniques, where first of all an edge detection is performed on the slices and afterwards the contours are connected. Having the contours determined the corresponding contour points in the neighboring slices are connected by triangles. This approach requires the setting of many heuristic parameters thus it is not flexible enough to use them in practical applications. A more robust approach is the “marching cubes” isosurface reconstruction \cite{Lorensen}, which marches through all the cubic cells and generates an elementary triangular mesh whenever a cell is found which is intersected by an iso-surface. Since the volumetric data defined in the discrete space is converted to a continuous geometrical model, the conventional computer graphics techniques, like ray tracing or buffering can be used to render the iso-surfaces.

Another indirect volume-rendering approach is known as 3D Fourier transform (3D FT), where the intermediate representation is a 3D Fourier transform of the volume rather than a geometrical model  \cite{Malzbender} \cite{Lippert} \cite{Totsuka}. This technique aims at fast density integral calculation along the viewing rays. Since the final image is considered to be an X-ray simulation, this technique is useful in medical imaging applications. The main idea is to calculate the 3D Fourier transform of the volume in a preprocessing step. This transformation is rather expensive computationally but it has to be executed only once independently on the viewing direction. The final image is calculated performing a relatively cheap 2D inverse Fourier transformation on a slice in the frequency domain. This slice is perpendicular to the current viewing direction and passes through the origin of the coordinate system. According to the Fourier projection-slice theorem the pixels of the generated image represent the density integrals along the corresponding viewing rays.

\subsubsection{Space domian volume rendering: Marching Cubes}

Isosurface is an operation that given a scene outputs a connected surface as a binary shell. Connectedness means that within the output shell it is possible to reach to any shell element from any shell element without leaving the shell. If the input is a binary scene, the shell constitutes a connected interface between 1-cells and 0-cells. If the input is a grey scene, the interface between the interior and exterior of the structure is usually difficult to determine. Thresholding can be used to determine this interface, in which the shell constitutes essentially a connected interface between cells that satisfy the threshold criterion and cells that do not. In a particular thresholding operation specified by a single intensity value, the resulting surface is called an iso-surface. The common iso-surfacing algorithms are Opaque Cubes (Cuberille) \cite{Herman} \cite{Roberts}, Marching Cubes, Marching Tetrahedra \cite{Cignoni}, and Dividing Cubes \cite{Borouchaki}.  Of which, the most popular one used for medical visualization today is Marching Cubes.

Marching Cubes algorithm, developed by Lorensen and Cline in 1987 \cite{Lorensen}  is used to approximate an isosurface by subdividing a region of space into 3D array of rectangular cells, which is the most popular method for isosurface rendering. Another popular isosurface extraction method  is a propagation-based marching cubes method in 1986 by Wyvill et al.\cite{Wyvill}. That method is somewhat similar to Marching cubes, yet they have some shortcomings. The isosurfaces they extract also differ. Due to the differences, and since most teams who have described application of a marching cube, in the report we restrict the Marching cubes designation to the Lorensen's approach. 

The basic idea of Marching Cubes is that voxel could be defined by the pixel values at the eight corners of the cube. If one or more pixels of a cube have values less than the user-specified isovalue, and one or more have values greater than this value, we know the voxel must contribute some component of the isosurface. By determining which edges of the cube are intersected by the isosurface, we can create triangular patches which divide the cube between regions within the isosurface and regions outside. By connecting the patches from all cubes on the isosurface boundary, we get a surface representation.

In the eighties the volume-rendering research was mainly oriented to the development of in- direct methods. At that time no rendering technique was available which could visualize the volumetric data directly without performing any preprocessing. The existing computer graphics methods, like ray tracing or z-buffering \cite{Szirmay-Kalos} had been developed for geometrical models rather than for volume data sets. Therefore, the idea of converting the volume defined in a discrete space into a geometrical representation seemed to be quite obvious. The early surface reconstruction methods were based on the traditional image-processing techniques \cite{Artzy}  \cite{Trivedi}  \cite{Udupa} , like edge detection and contour connection. Because of the heuristic parameters to be set these methods were not flexible enough for practical applications, they are lacking detail and introducing artifacts.  Lorensen and Cline \cite{Lorensen} came up with the idea of creating polygonal representation of constant density surfaces from 3D array of data. Existing methods of 3D surface generation by Wyvill et al.\cite{Wyvill} trace contours within each slice then connect with triangles ( topography map), create surfaces from “cuberilles” (voxels), perform ray casting to find the 3D surface using hue-lightness to shade surface and gradient to shade, and then display density volumes.  There are some shortcomings of Wywill et al's techniques. One thing is that they throw away useful information in the original data, in cuberilles level they use thresholding to represent surface, during the process of ray casting, they use depth shading alone or approximates shading using unnormalized gradient. Another thing is that these methods lack hidden surface removal, and volume models display all values and rely on motion to produce a 3D sensation.  Thus Marching Cubes algorithm is introduced, for Marching Cubes algorithm uses all information from source data, derives inter-slice connectivity, surface location, and surface gradient, also result of Marching Cubes can be displayed on conventional graphics display systems using standard rendering algorithms, and also does not rely on image processing performed on the slices and requires only one parameter which is a density threshold defining the isosurface. 

In summary, marching cubes creates a surface from a three-dimensional set of data as follows \cite{Lorensen}:

\begin{enumerate}
  \item  Read four slices into memory;
  \item  Scan two slices and create a cube from four neighbors on one slice and four neighbors on the next slice;
  \item  Calculate an index for the cube by comparing the eight density values at the cube vertices with the surface constant;
  \item  Using the index, look up the list of edges from a precal- culated table;
  \item  Using the densities at each edge vertex, find the surface and edge intersection via linear interpolation;
  \item  Calculate a unit normal at each cube vertex using central differences. Interpolate the normal to each triangle vertex;
  \item  Output the triangle vertices and vertex normals. 
\end{enumerate}

\begin{figure}[h]  
\centering
    \includegraphics[width=0.5\textwidth]{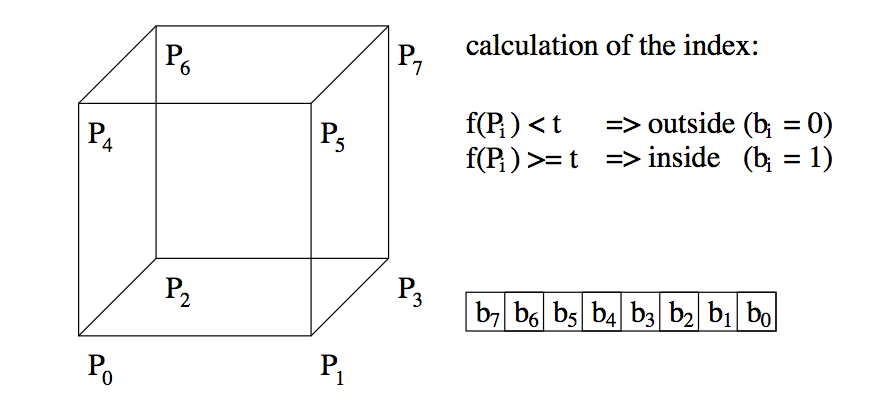}
    \caption{Calculation of the index to the look-up table}
    \label{fig:MC_lookuptable}
\end{figure}

After having an isosurface defined by a density threshold, all the voxels are investigated whether they are below or above the surface, comparing the densities with the surface constant. To locate the surface, it uses a logical cube created from eight pixels, 4 each from 2 adjacent layers, slice $k$ and slice $k+1$. This binary classification assigns the value of one to the voxels of densities higher than the threshold and the value of zero to the other voxels, which sets cube vertex to value of 1 if the data value at that vertex exceeds or equals the value of the surface we are constructing
otherwise, sets cube vertex to 0. If a vertex  = 1 then it is “inside” the surface, if a vertex  = 0 then it is “outside”. Any cube with vertices of both types is “intersected” by the surface. The algorithm marches through all the intersected cells, where there are at least two corner voxels classified differently. For such cells an index to a look-up table is calculated according to the classification of the corner voxels as shown in Figure~\ref{fig:MC_lookuptable} .

\begin{figure}[h]  
\centering
    \includegraphics[width=0.5\textwidth]{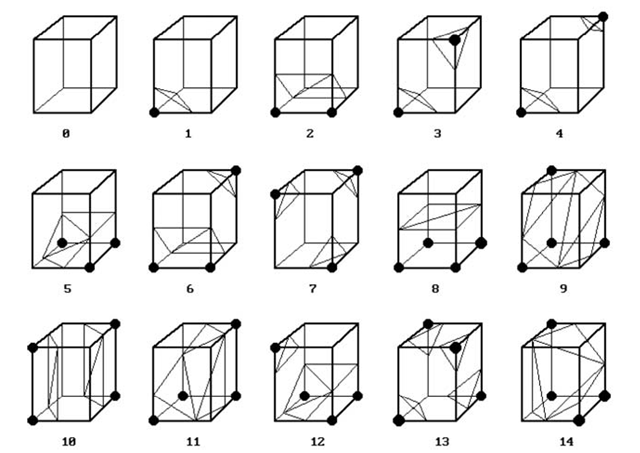}
    \caption{Marching Cubes}
    \label{fig:MC}
\end{figure}

For each cube, we have 8 vertices with 2 possible states each, inside or outside. This gives us $2^{8}$ possible patterns, which is 256 cases. An index system is built, which contains eight bits associated with the eight corner voxels of the cubic cell and their values depend on the classification of the corresponding voxels. This index addresses a look up table containing all the 256 cases of elementary triangular meshes. Because of symmetry reasons, there are just 15 topologically distinct cases among these patterns thus in practice the look-up table contains 15 entries instead of 256. Figure~\ref{fig:MC} shows the triangulation of the 15 patterns.

After having the list of intersected edges read from the look up table,  the intersection points along these edges are calculated using linear interpolation. Vertex bit mask is used to create an index for each case based on the state of the vertexes, and then index will tell which edge the surface intersects, we can then can linearly interpolate the surface intersection along the edge.
The position of vertex $V_{i,j}$ along the edge connecting corner points $P_{i}$ and  $P_{j}$ is computed as  \eqref{eq:V(i,j)}:

\begin{equation}
V_{i,j}=((t-f(P_{i})*P_{j}))+\frac{(f(P_{j})-t)*P_{i}}{f(P_{j})-P_{i}}   \label{eq:V(i,j)}
\end{equation}

assuming that $ f(P_{i})<t $ and $ f(P_{i})>t $, where $f$ is the spatial density function and $t$ is the threshold defining the isosurface.

Since the algorithm generates an oriented surface with a normal vector at each vertex position, the last step is the calculation of the surface normals.

To calculate surface normal, we need to determine gradient vector, $G$ , which is the derivative of the density function. To estimate the gradient vector at the surface of interest, we first estimate the gradient vectors at the vertices and interpolate the gradient at the intersection.
The gradient at cube $V_{i , j, k} $, is estimated using central differences along the three coordinate axes by:

\begin{equation}
G_{i,j,k}(x)=\frac{f(x_{i+1}, y_{j}, z_{k})-f(x_{i-1}, y_{j}, z_{k})}{\delta x}
\end{equation}
\begin{equation}
G_{i,j,k}(y)=\frac{f(x_{i}, y_{j+1}, z_{k})-f(x_{i}, y_{j-1}, z_{k})}{\delta y}
\end{equation}
\begin{equation}
G_{i,j,k}(z)=\frac{f(x_{i}, y_{j}, z_{k+1})-f(x_{i}, y_{j}, z_{k-1})}{\delta z}
\end{equation}

The normals $ n(x_{i}, y_{j}, z_{k}) $
at the cube vertices are determined using central differences  \eqref{eq:n(x_{i}, y_{j}, z_{k})}:

\begin{equation}
n(x_{i}, y_{j}, z_{k})\approx \frac{1}{2}*\begin{bmatrix}
f(x_{i+1}, y_{j}, z_{k})-f(x_{i-1}, y_{j}, z_{k})\\ 
f(x_{i}, y_{j+1}, z_{k})-f(x_{i}, y_{j-1}, z_{k})\\ 
f(x_{i}, y_{j}, z_{k+1})-f(x_{i}, y_{j}, z_{k-1})
 \label{eq:n(x_{i}, y_{j}, z_{k})}
\end{bmatrix}
\end{equation}

After dividing the gradient by its length produces the unit normal at the vertex required for rendering.
Then the algorithm linearly interpolates this normal to the point of intersection. At an intersection point  $V_{i,j}$ along the edge connecting grid points $P_{i}$ and $P_{j}$ the surface normal $ N_{i,j} $ is calculated using linear interpolation between the corresponding normals denoted by $ N_{i} $  and $ N_{j} $ respectively \eqref{eq:N(i,j)}:

\begin{equation}
N_{i,j}=((t-f(P_{i})*N_{j}))+\frac{(f(P_{j})-t)*N_{i}}{f(P_{j})-P_{i}}   \label{eq:N(i,j)}
\end{equation}

The continuous geometrical model generated by the marching cubes algorithm can be ren- dered using the traditional computer graphics techniques. The conventional graphics accelera- tion devices which are based on the hardware implementation of the -buffering hidden surface removal can render such a model in real time using Phong shading or Gouraud shading. The pseudocode of generalized Marching Cubes algorithm is shown as following:

 Pseudocode: Generalized Marching Cubes \cite{Lorensen}
\begin{lstlisting}[language={[ANSI]C}, basicstyle=\ttfamily\tiny, numbers=left, numberstyle=\tiny, keywordstyle=\color{blue!70}, commentstyle=\color{red!50!green!50!blue!50}, frame=shadowbox, showstringspaces=false, rulesepcolor=\color{red!20!green!20!blue!20}, xleftmargin=2em,xrightmargin=0em, aboveskip=1em]
Create an edge table
Read in 3 2d slices of data
while(moreDataSlices)
 {
 Read in next slice of data
 while(moreCubes) 
  {
  Fill cube index 
  Assign 1 or 0 to vertex index
  edgesToDrawBetween = edgeTable[cubeIndex]
  Interpolate:
       triangle vertexes from edge vertexes
  Determine triangle vertex normals
  Draw triangle(s)
  currentCube++;
 }
 Discard data slice 
}
\end{lstlisting}

Marching cubes algorithm has been applied in many application areas, including biochemistry \cite{Heiden}, biomedicine \cite{Yim}, deformable modeling \cite{Lin}, digital sculpting \cite{Ferley}, environmental science \cite{Stein}, mechanics and dynamics \cite{Matsuda}, natural phenomena rendering \cite{Trembilski}, visualization algorithm analysis \cite{Kim}, etc. Processing involving depth maps \cite{Savarese} has also been influenced by Marching cubes isosurface extraction, especially in the development of methods based on distance fields \cite{Frisken}.

\begin{figure}[h]  
\centering
    \includegraphics[width=0.5\textwidth]{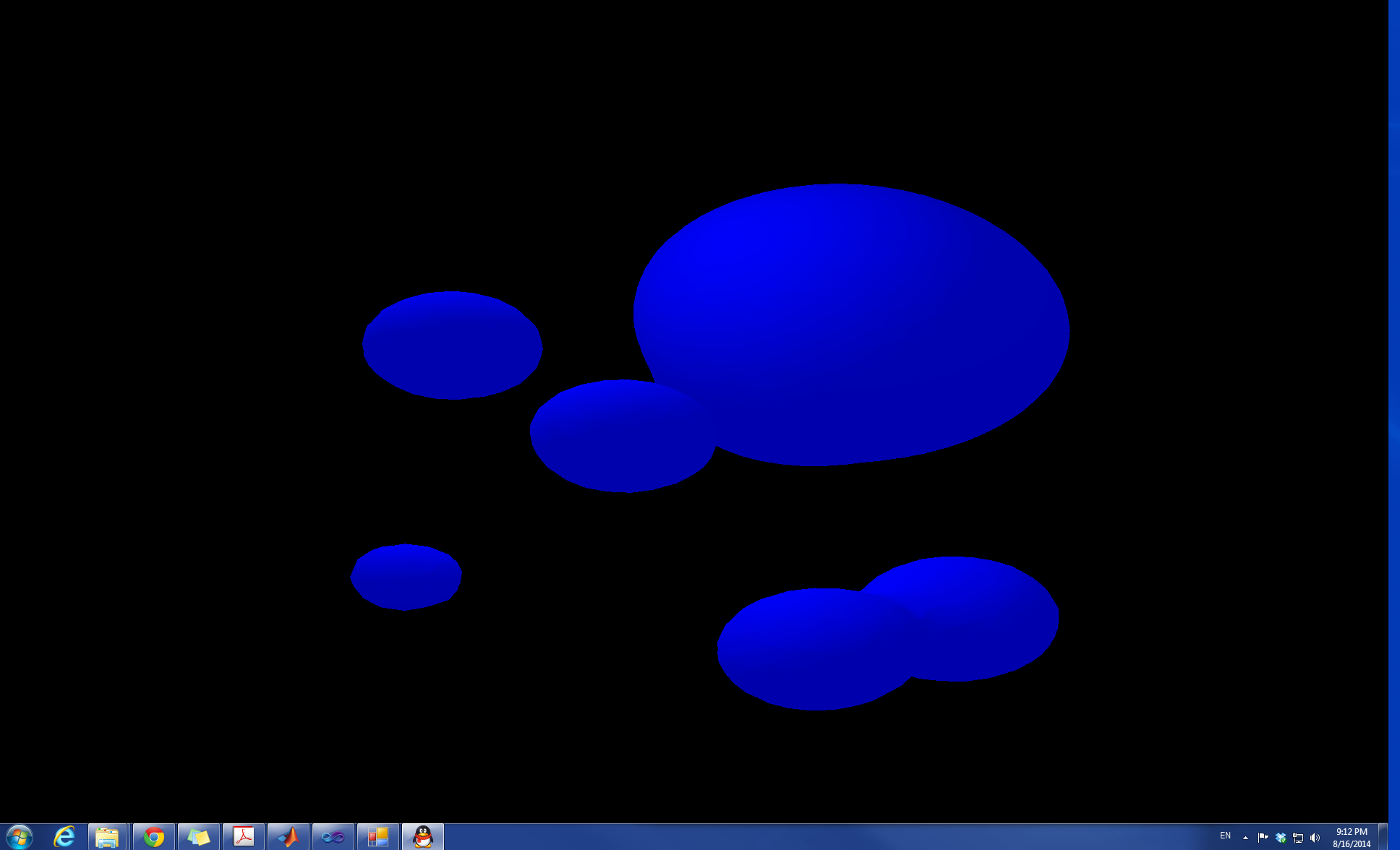}
    \caption{Marching Cubes of Metaballs}
    \label{fig:MC2}
\end{figure}

An preliminary implementation of marching cubes is shown as below Figure~\ref{fig:MC2} .

Although employed in many arenas, isosurface creation is heavily utilized in medical visualization \cite{Dedual} and computer-aided diagnosis applications \cite{Meng-fei}. Isosurfaces recreate the digitized images taken by computed tomography (CT), magnetic resonance (MR), and single-photon emission computed tomography(SPECT).

Pseudocode: Marching Cube for Medical Image Dataset \cite{Delibasis}
\begin{lstlisting}[language={[ANSI]C}, basicstyle=\ttfamily\tiny, numbers=left, numberstyle=\tiny, keywordstyle=\color{blue!70}, commentstyle=\color{red!50!green!50!blue!50}, frame=shadowbox, showstringspaces=false, rulesepcolor=\color{red!20!green!20!blue!20}, xleftmargin=2em,xrightmargin=0em, aboveskip=1em]
for each image voxel
   a cube of length 1 is placed on 
   eight adjacent voxels of the image
   for each of the cube's edge{
        if (the one of the node voxels has value 
            greater than or equal to t and 
            the other voxel has value less than t) 
        then
            {calculate the position of a point on
             the cube's edge that belongs to the 
             isosurface, 
             using linear interpolation}
            }
    for each predefined cube configurations{
        for each of eight possible rotations{
             for the configuration's complement{
                compare the produced cube configuration 
                of the above calculated isopoints to the 
                set of predefined cube configurations 
                and produce the corresponding triangles
             }
        }
     }
}

\end{lstlisting}

Marching Cubes algorithm is capable only for isosurface rendering thus the internal structures of the volume cannot be visualized. After the preprocessing, the original data values are not available anymore thus cutting planes are not supported. Cutting operations are rather important in medical imaging applications, where the physician can define an arbitrary cross section of the 3D model and render the slice displaying the original gray scale data values. Furthermore, the modeling of semi-transparent tissues, which is the most important feature of direct volume rendering,  is not supported either.

The main disadvantage of the marching cubes algorithm is the computationally expensive preprocessing. Especially having high resolution data sets the number of the generated trian- gles can be enormous. Since the interactivity is strongly influenced by the complexity of the model usually some postprocessing is performed on the initial mesh in order to simplify it \cite{Shekhar}. Furthermore the triangular mesh is not uniform because the vertices are located on the edges of cubic cells, therefore some mesh refinement is also required.

Advantages of Marching Cubes: 

\let\oldenumerate\enumerate
\renewcommand{\enumerate}{
  \oldenumerate
  \setlength{\itemsep}{0.1pt}
  \setlength{\parskip}{0pt}
  \setlength{\parsep}{0pt}
}

\begin{enumerate}
 \item Uses all information from source data;
  \item Derives inter-slice connectivity, surface location, and surface gradient;
  \item Result can be displayed on conventional graphics display systems using standard rendering algorithms; 
  \item Allows Solid modeling capability: cutting and capping.
\end{enumerate}

Disadvantages of Marching Cubes:

\begin{enumerate}
  \item Requires user input; 
  \item There is a loss of accuracy when visualizing small or fuzzy details;
  \item The assumptions which are made about the data may not necessarily be valid. This particularly applies to the assumption that the surfaces exist within the data to map the geometric primitives onto;
  \item Unless the original information is stored along with the geometric representation, the information on the interior of the surfaces is lost;
  \item Mainly limited to medical images with clear contiguous intensity boundaries: constant density; 
  \item Is performing a modified form of thresholding.
\end{enumerate}

\subsubsection{Frequency domain volume rendering:Fourier Transform}

Fourier Volume Rendering (FVR) developed by Levoy, Totsuka and Malzbender \cite{Totsuka} is based on the frequency spectrum of the 3D scalar field by utilizing the Fourier Slice Projection theorem  \cite{Lim}  \cite{Oppenheim}. This theorem allows us to compute integrals over volumes by extracting slices from the frequency domain representation. It states that a projection of a 3D data volume from a certain view direction can be obtained by extracting a 2D slice perpendicular to that view direction out of the 3D Fourier spectrum and then inverse Fourier transforming it \cite{Levoy}, as is show in Figure~\ref{fig:FTV} :

\begin{figure}[h]  
\centering
    \includegraphics[width=0.4\textwidth]{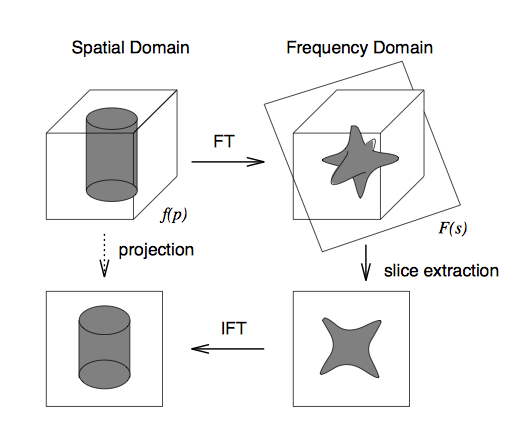}
    \caption{Frequency domain rendering}
    \label{fig:FTV}
\end{figure} 

Malzbender proposes various filters for high-quality resampling in frequency domain. Totsuka and Levoy \cite{Totsuka} extended this work with depth cues and shading performing calculations in the frequency domain during slice extraction. 

Illumination models for FVR were studied in the work of \cite{ENTEZARI}. They describe methods to integrate diffuse lighting into FVR. One approach is based on gamma corrected hemispherical shading and is suitable for interactive rendering of fixed light sources. Another technique uses spherical harmonic functions and allows lighting using varying light sources. These shading techniques, however, require a large amount of memory and are not well suited for visualization of large data sets. Another approach that produces images which are similar to FVR is based on importance sampling and Monte Carlo integration \cite{Laszlo} thus the samples are not aligned ona regular grid. This technique overcomes the limitation of parallel projection and the overall computational complexity $O(N^2)$  is better than in case of FVR.

Fourier Slice Projection theorem, also known as Fourier slice theorem or projection slice theorem in mathmatics, is a theorem states that the results of the following two calculations are equal. The first calculation procedure is stated as,  take a three dimensional function $f(r)$ project it onto a two dimensional plane, and do a Fourier transform of that projection. The second calculation procedure is stated as take that same function, but do a three dimensional Fourier transform first, and then slice it through its origin, which is parallel to the projection line.

For a 3D volume, the theorem states that the following two are a Fourier transform pair, the 2D image obtained by taking line integrals of the volume along rays perpendicular to the image plane,
and the 2D spectrum obtained by extracting a slice from the Fourier transform of the volume along a plane that includes the origin and is parallel to the image plane. 

The inverse process of Fourier Volume Rendering can be seen as a reconstruction method. Here, a set of pre acquired projections are Fourier  transformed and then put slice by slice into a Fourier Volume, which is initialized with zeros. If enough projections are available the Fourier Volume will be filled completely after a while. Consequently, by applying the inverse Fourier trans- form to the reconstructed Fourier Volume the spatial representation of the object, described by the projections, can be computed.

The general pipeline of the Fourier Volume Rendering technique can be devided into two basic steps. At first a computationally expensive one-time preprocessing step is performed. In a second rendering step arbitrary view directions can then quickly be computed by carrying out a two dimensional slicing operation and an inverse frequency transform.

\begin{figure}[h]  
\centering
    \includegraphics[width=0.4\textwidth]{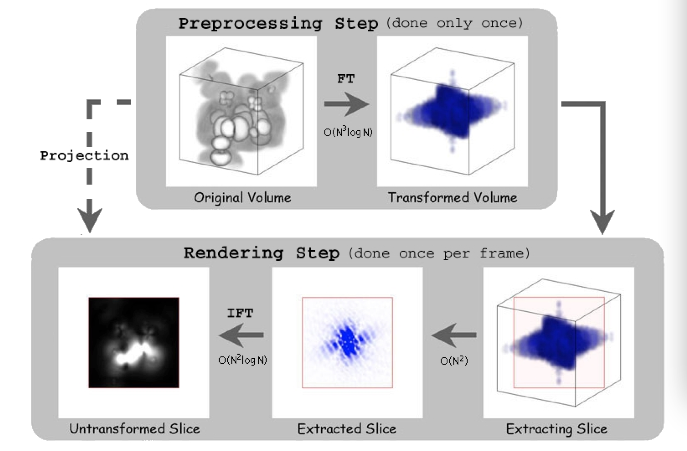}
    \caption{Fourier Volume Rendering pipeline}
    \label{fig:FVRpipeline}
\end{figure} 

Figure~\ref{fig:FVRpipeline} gives a more detailed overview of the involved operations. Here, the preprocessing step transforms the spatial domain volume into its frequency domain representation. This is usually accomplished with either the three dimensional Fast Fourier Transform or the Fast Hartley Transform and has a run time complexity of   $O(N^3 log N)$  assuming an $N*N*N$ input data set. Due to the Fourier Projection Slice Theorem a view from any arbitrary angle can then easily be computed in a second rendering step by slicing the frequency volume along a plane oriented perpendicular to the viewing direction and crossing exactly through the center of the frequency volume. On account of the two dimensional character of the resultant slice, the overall complexity of this operation is $O(N^2)$. Finally the sought-after projection is derived by taking an inverse two-dimensional Fast Fourier or Fast Hartley Transform of the frequency slice. The complexity of this operation is   $O(N^2 log N)$, which is the asymptotic running time of the rendering step.

In summary, FVR algorithm works as following:
\begin{enumerate}
\item Load the volume data into memory and compute its 3D Fourier transform using the FFT algorithm; 
\item When the view direction changes, take a 2D slice of the volume, passing through the its center and perpendicular to the view direction. Many operations can be performed, at this stage, on the 2D slice to get various effects such as thresholding, depth cue, etc, through the use of filters;
\item Take the 2D inverse Fourier transform of this slice. Rescale its values to valid intensity range and display the slice.
\end{enumerate}

A straightforward implementation of the Fourier transform is not suitable for high performance FVR. The inverse two dimensional transform must be computed at high speed to achieve interactive framerates. Therefore fast variants of the Fourier Transform are used in FVR implementations. The original idea of the Fast Fourier Transform (FFT) was introduced by Cooley \cite{COOLEY}. Their algorithm decomposes the Discrete Fourier Transform (DFT) into $log(2N)$ passes, where $N$ is the size of the input array. Each of these passes consists of $N/2$ butterfly computations. Each butterfly operation takes two complex numbers a and b and computes two numbers, $a+wb$  and $a-wb$, where $w$ is a complex number, called principal $N$th root of unity.  After $log(2N)$  passes the butterfly operations result into the transformed data. One of the fastest implementations available, is the FFTW library \cite{Frigo}.

The Fast Hartley Transform (FHT) \cite{Bracewell_1} performs as an alternative to FFT. The transform produces real output for a real input, and is its own inverse. Therefore for FVR the FHT is more efficient in terms of memory consumption. The Multidimensional Hartley Transform, however, is not separable. The N  dimensional transform cannot be computed as a product of N one dimensional transforms. Bracewell and Hao propose a solution to this problem \cite{Bracewell_1}  \cite{Bracewell_2}. They suggest to perform N one dimensional transformations in each orthogonal di- rection followed by an additional pass that corrects the result to correspond to the N dimensional Hartley transform.

\begin{figure}[h]  
\centering
    \includegraphics[width=0.4\textwidth]{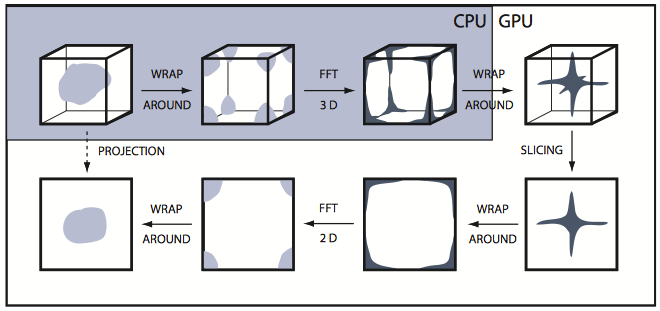}
    \caption{Frequency domain volume rendering pipeline on GPU}
    \label{fig:FVR_GPU}
\end{figure} 

Based on the work of Bracewell and Hartley, a GPU based FVR algorithm \cite{Ivan}, and which is accelerated by factor of 17 by mapping the rendering stage to the GPU.  The three dimensional transformation into frequency domain is done in a preprocessing step. The rendering step is computed completely on the GPU. First the projection slice is extracted. Four different interpolation schemes are used for resampling the slice from the data represented by a 3D texture. The extracted slice is trans- formed back into the spatial domain using the inverse Fast Fourier or Fast Hartley Transform. The rendering stage is implemented through shader programs running on programmable graphics hardware achieving highly interactive framerates, and the pipeline is as shown in Figure~\ref{fig:FVR_GPU}.

Pseudocode: Fourier Volume Rendering on GPU \cite{Ivan}
\begin{lstlisting}[language={[ANSI]C}, basicstyle=\ttfamily\tiny, numbers=left, numberstyle=\tiny, keywordstyle=\color{blue!70}, commentstyle=\color{red!50!green!50!blue!50}, frame=shadowbox, showstringspaces=false, rulesepcolor=\color{red!20!green!20!blue!20}, xleftmargin=2em,xrightmargin=0em, aboveskip=1em]
/* Initializing necessary variables */ 
InitVars(); 
/* Initialize CUDA context (GL context has been created already) */ 
InitCUDAContext(argc, argv, true);

/* Reading & initializing volume data */ 	
InitData(); 

/* Creating float volume & releasing byte data */ 	
CreateFloatData();

/* CUDA or Hybrid pipeline */
if (!GPU)
{
  /* Wrapping around spatial volume */  
   WrapAroundVolume(); 
  /* Creating spectrum complex arrays */ 
  CreateSpectrum(); 
  /* Wrapping around spectral volume */  
  WrapAroundSpectrum(); 
  /* Packing spectrum complex data into texture  
   * array to be sent to OpenGL */ 
   PackingSpectrumTexture(); 	
}
else 
{
  /* Spectral texture for OpenGL compatability */
  mTextureArray = 
   (float*) malloc (mVolumeSize * 2 * sizeof(float)); 
  /* Run the FVR on the CUDA pipeine */
  CUDA_Way();
}

/* Uploading spectrum texture to GPU for slicing */
SendSpectrumTextureToGPU(); 
/* We don't need float data ayn more as it resides in the 
 * GPU texture memory */ 	
delete [] mVolumeDataFloat; 

/* Intersecting QUAD with the texture */  
SetDisplayList();

/* CUDA timer */ 
cutCreateTimer(&mTimer);
cutResetTimer(mTimer);

/* Register OpenGL callbacks */ 
glutDisplayFunc(DisplayGL);
glutKeyboardFunc(KeyBoardGL);
glutReshapeFunc(ReshapeGL);
glutIdleFunc(IdleGL);

/* Initializing OpenGL buffers */ 
InitOpenGLBuffers();

/* Start main rendering loop */ 
glutMainLoop();

/* Clean Up */ 
CleanUp(EXIT_FAILURE);

/* Exiting ... */ 
// shrEXIT(argc, (const char**)argv);
\end{lstlisting}

Advantages of Fourier Volume Rendering:

\begin{enumerate}
 \item  Big improvement in speed;\\
 \item  Permits the rendering of compressed datasets by using only portions of the frequency spectrum. Fourier Volume Rendering also allows the application of lowpass, highpass, and bandpass filters with little overhead, since the volume data are already available in an adequate representation. By exploiting this property other operations such as successive refinement can easily be achieved just by successively adding higher and higher frequencies to the resultant image. This also permits the rendering of compressed datasets by using only portions of the frequency spectrum, which might be useful for web applications.
\end{enumerate}

Disadvantages of Fourier Volume Rendering:

\begin{enumerate}
  \item   Lack of occlusion and hidden surfaces, the projection obtained by the Fourier projection slice theorem is a line integral normal to the direction of view. Voxels on a viewing ray contribute equally to the image regardless of their distance from the eye. The image therefore lacks occlusion, an important visual cue. While some users prefer integral projections since nothing is hidden from view, this characteristic would be considered a drawback in most applications. Since for the calculation of the density integrals a distance-dependent weighting function or an opacity manipulation can- not be used. In the next section it will be shown that direct volume-rendering methods are much more flexible in this sense supporting the modeling of several optical phenomena like emission, reflection, and attenuation.\\
 \item  Exclusive support of orthogonal viewing;\\
 \item  Contrast of high interpolation costs versus ghosts;\\
 \item  Significantly higher memory costs.
\end{enumerate}

All problems are technical in nature and several solutions are proposed, yet the lack of occlusion is fundamental and so far no projection slice theorem is known that reproduces the integral differential equation approximated by volume rendering algorithms.

\subsection{Direct volume rendering}

Direct volume rendering is a method which renders the data set directly without using any intermediate representation. The optical attributes like a color, an opacity, or an emission are assigned directly to the voxels. The pixel colors depend on the optical properties of the voxels intersected by the corresponding viewing rays.

In comparison to the indirect methods presented in the previous section, direct methods display the voxel data by solving the equation of radiative transfer for the entire volumetric object. In direct volume rendering, the scalar value given at a sample point is virtually mapped to physical quantities that describe the emission and absorption of light at that point. This mapping is also often termed classification. It is usually performed by means of a transfer function that maps data values to color (emission) and opacity (absorption). These quantities are then used for a physically based synthesis of virtual images.

Similar to a divide and conquer-strategy, algorithms for direct volume rendering differ in the way the complex problem of image generation is split up into several subtasks. A common classification scheme differentiates between image order and object order algorithms. The direct volume rendering pipeline is shown in Figure~\ref{fig:DirectPipeline} :

\begin{figure}[h]  
\centering
    \includegraphics[width=0.4\textwidth]{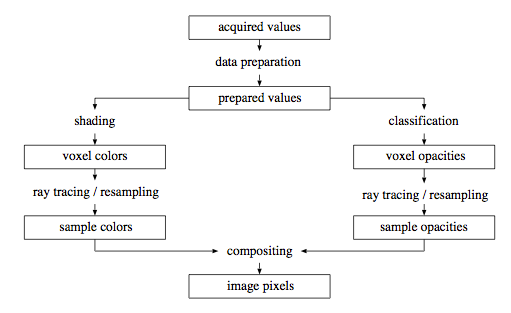}
    \caption{The Direct Volume Rendering Pipeline}
    \label{fig:DirectPipeline}
\end{figure} 

The first step in direct volume rendering is data preparation, before any render pipeline operation is performed on the data, it may need some sort of preparation first. Filtering, anti aliasing, contrast enhancement, domain switching are some common operations that are used.  he prepared array is the input of the shading process where colors are assigned to the voxels depending on their densities. The assigned colors are shaded. The shading model requires a normal vector at each voxel location. In gray level volumes the normals can be obtained as the estimated gradients calculated from the central differences as shown below, 

\begin{equation}
\bigtriangledown f(x_{i}, y_{j}, z_{k})\approx \frac{1}{2}\begin{bmatrix}
f(x_{i+1}, y_{j}, z_{k})-f(x_{i-1}, y_{j}, z_{k})\\ 
f(x_{i}, y_{j+1}, z_{k})-f(x_{i}, y_{j-1}, z_{k})\\ 
f(x_{i}, y_{j}, z_{k+1})-f(x_{i}, y_{j}, z_{k-1})
\end{bmatrix}
\end{equation}

where $f(x_{i}, y_{j}, z_{k})$ is the discrete 3D density function. The output of the shading process is an array of voxel colors. In a separate step, classification is performed yielding an additional array of voxel opacities. After having the color and opacity values assigned to each voxel, rays are cast from the view point (image order) or object point (object order) and perform the process of resampling the volume. The treated samples are rendered to screen in the final step. A number of optical and illumination models can be used depending on how realistic the final image has to be or how computationally complex the operation is allowed to be. 

The simplest visualization models directly map the density profile onto pixel intensities. For instance, one possibility is to calculate each pixel value $I(x,w)$ as the density integral along the corresponding viewing ray defined by origin $x$ and direction $w$ :

\begin{equation}
I(x,w)=\int_{t}^{} f(x+w*t) dt
\end{equation}

This model is equivalent with the Fourier volume rendering resulting in simulated X-ray im- ages. Similar visual effect can be achieved approximating the density integrals by the maximum density value along the viewing rays, as below: 

\begin{equation}
I(x,w)=max_{t} f(x+w*t)
\end{equation}

There are several models available as shown in Figure~\ref{fig:RayTraversalSchemes} :

\begin{figure}[h]  
\centering
    \includegraphics[width=0.4\textwidth]{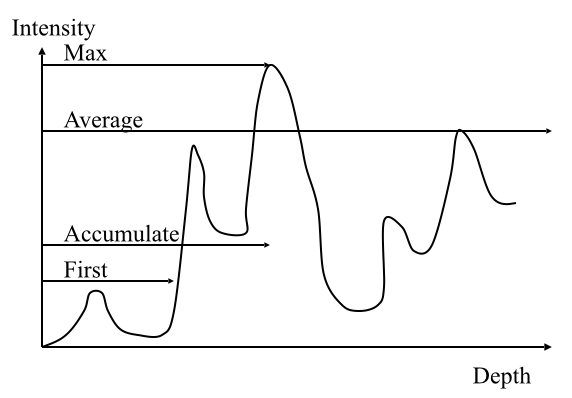}
    \caption{Ray Traversal Schemes}
    \label{fig:RayTraversalSchemes}
\end{figure} 

For early methods of direct volume rendering, transparency was not considered till 1988, researchers did not consider sophisticated light transportation theory, and were concerned with quick solutions, hence models at that time were more or less applied to binary data, since non binary data requires sophisticated classification and compositing methods.  The first ray traversal was only considering the first volume data the ray reaches for extracting the iso surface, it was developed by Tuy in 1984 \cite{Tuy1984} .

And then researchers came up with the idea of taking  average value ray traversed for representation,  and this method produces basically an X-ray picture.  Some researchers came up with another idea that they treated Maximum Intensity Projection for magnetic resonance image (MRI) rendering.  Accumulate opacity method was developed by Levoy in 1988, while compositing colors this method makes transparent layers visible. \cite{Levoy1988}

Among all models introduced by tons of researchers, one well known in medical imaging as maximum intensity projection (MIP) and it is mainly used for visualization of blood vessel structures. Assuming that a trilinear filter is applied for function reconstruction the exact maximum densities can be analytically calculated \cite{Sakas1995} . In practice the density profile is approximated by a a piecewise constant function taking a finite number of evenly located samples, and the maximum density sample is assigned to the given pixel.

The main drawback of maximum intensity projection is the loss of depth information. For example, in a medical application it might be confusing that a higher density blood vessel can “hide” other blood vessels which are closer to the view-point. In order to avoid this problem Sato proposed a technique called local maximum intensity projection (LMIP) \cite{Sato1998} . Instead of the global maximum along the corresponding viewing ray the first local maximum which is above a predefined threshold is assigned to each pixel, as is shown below Figure~\ref{fig:LMIP} :

\begin{figure}[h]  
\centering
    \includegraphics[width=0.4\textwidth]{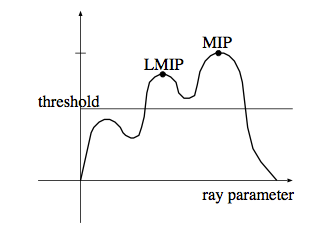}
    \caption{Local Maximum Intensity Projection }
    \label{fig:LMIP}
\end{figure}

In order to model physical phenomena like scattering or attenuation optical properties are assigned to the volume samples as functions of their density values. Each pixel intensity $I(x,w)$
is composed from the assigned properties of the samples along the associated viewing ray according to the well known light transport equation \cite{Levoy1988} \cite{Levoy1990} \cite{Danskin1992} \cite{Max1995}:

\begin{equation}
I(x,w)=\int_{s} e^{\int_{0}^{s}\sigma(t)dt}I(s)ds
\end{equation}

or in its discrete form, as in the limit as the sample spacing goes to zero, is approximated by a summation of equally spaced samples \cite{Max1990} \cite{Sabella1988} :

\begin{equation}
I=\sum_{0\leqslant i \leqslant n}( \prod_{0\leqslant i \leqslant j }e^{-\int_{j}^{j+1}\sigma (t)dt})I(i)
\end{equation}

where $x$ is the origin of the ray, $w$ is the unit direction of the ray, $\sigma(t)$ s the differential attenuation at $x+w*t$, and $I(s)$ is the differential intensity scattered at $x+w*s$ in the direction $-w$.

Introducing $\alpha (j)$  as the accumulated opacity of ray segment $[j,j+1]$, this method is also known as alpha blending :

\begin{equation}
\alpha (j)=1-e^{-\int_{j}^{j+1}\sigma (t)dt}
\end{equation}

$I$ can be evaluated recursively running through each $i th$ sample in back to front order:

\begin{equation}
I_{out}=\alpha (i)*I(i)+(1-\alpha (i))*I_{in}
\end{equation}

where $I_{in}$ is the intensity before the evaluation of the $i th$ sample, $I(i)$ is the scattering of the  $i th$ sample, and  $I_{out}$ is the intensity after having the contribution of the $i th$ sample added to the weighted sum. The initial value of  $I_{in}$ is the ambient light. In fact, $I_{out}=\alpha (i)*I(i)+(1-\alpha (i))*I_{in}$ is the Porter Duff over operator used for compositing digital images \cite{Porter1984} . In the following subsections two different strategies are presented for approximating the light transport equation using the over operator.

Pseudocode:  Alpha Blending
\begin{lstlisting}[language={[ANSI]C}, basicstyle=\ttfamily\tiny, numbers=left, numberstyle=\tiny, keywordstyle=\color{blue!70}, commentstyle=\color{red!50!green!50!blue!50}, frame=shadowbox, showstringspaces=false, rulesepcolor=\color{red!20!green!20!blue!20}, xleftmargin=2em,xrightmargin=0em, aboveskip=1em]
rgb AlphaBlending(int x, int z) { 
int y, i, segment;
rgb color = BLACK;
voxel v;
double trans = 1.0; for(y = 0; y < 8; y++) {
    segment = mask[z][y][x]);
    while(segment) {
i = Trace(segment);
v = volume[z][y][x-offset[y*32+i];
 trans *= 1.0 - v.opacity;
if(trans < threshold) return color; 
color += v.color * v.opacity * trans; 
segment &=  ~(0x80000000 >> i);
}
 }
  return color;
}

\end{lstlisting}

In this report,  direct volume rendering techniques are classified further into two categories. The object order methods process the volume voxel by  voxel projecting them onto the image plane, while the image-order methods produce the image pixel by pixel casting rays through each pixel and resampling the volume along the viewing rays.

The direct techniques represent a very flexible and robust way of volume visualization. The internal structures of the volume can be rendered controlled by a transfer function which assigns different opacity and color values to the voxels according to the original data value. Although there is no need to generate an intermediate representation direct volume rendering is rather time consuming because of the enormous number of voxels to be processed.

\subsubsection{Image order methods}
Image order rendering, also called backward mapping, ray casting, pixel space projection, or image-space rendering,  is fundamentally different from object order rendering. Image order techniques consider each pixel of the resulting image separately. For each pixel, the contribution of the entire volume to this pixels’s final color is computed.

By the late 80's, a number of surface extraction techniques had been developed, like the marching cubes algorithm. Surface extraction and rendering using polygons works fairly well on arbitrary data, but it does have one big drawback; aliasing effects due to the difficulty of classifying where the actual surface is. This problem is further amplified by the fact that the marching cubes algorithm uses a binary classification scheme which means that a data point is either on the surface, or it is not. Obviously, that kind of classification is not very precise, and it does not work well even with real numbers. Polygons are generally not well suited to display, complex, fine details, especially since a near infinite number is needed. The classification errors produce aliasing artefacts, meaning that surface features that do not exist in the dataset are embedded in the final rendered picture. For applications used in the medical field in particular, aliasing or any artefacts at all are not acceptable.

One solution to the classification problem is to use a technique called raycasting. The basic algorithm is simple, rays are cast into a data volume and samples are taken along each ray by interpolation of the surrounding voxels. This means that no intermediate geometry is constructed and thus the classification problem is solved. Another advantage is that the volume can be rendered semi-transparent, and as a result it is possible to display many surfaces within each other.

Rene Descartes introduced ray tracing back in 1637,  the idea of tracing light rays and their interaction between surfaces. He applied the laws of refraction and reflection to a spherical water droplet to demonstrate the formation of rainbows. The first ray casting algorithm used for rendering was presented by Arthur Appel in 1968 \cite{Appel1968} . The idea behind ray casting is to shoot rays from the eye, one per pixel, and find the closest object blocking the path of that ray. Using the material properties and the effect of the lights in the scene, this algorithm can determine the shading of this object. The simplifying assumption is made that if a surface faces a light, the light will reach that surface and not be blocked or in shadow. The shading of the surface is computed using traditional 3D computer graphics shading models. One important advantage ray casting offered over older scanline algorithms is its ability to easily deal with non-planar surfaces and solids. If a mathematical surface can be intersected by a ray, it can be rendered using ray casting. Elaborate objects can be created by using solid modeling techniques and easily rendered.

In 1980, Turner Whitted \cite{Turner1980} used the basic ray casting algorithm but extended it. When a ray hits a surface, it could generate up to three new types of rays, reflection, refraction, and shadowing.  A reflected ray continues on in the mirror reflection direction from a shiny surface. It is then intersected with objects in the scene; the closest object it intersects is what will be seen in
the reflection. Refraction rays travelling through transparent material work similarly, with the addition that a refractive ray could be entering or exiting a material. To further avoid tracing all rays in a scene, a shadow ray is used to test if a surface is visible to a light. A ray hits a surface at some point. If the surface at this point faces a light, a ray is traced between this intersection point and the light.

Back in 1984, method cast parallel or perspective rays from the pixels of the image plane was proposed, Tuy's work known as binary ray casting determines only the first intersection points with a surface contained in the volume \cite{Tuy1984}.  Binary ray casting aims at the visualization of surfaces contained in binary volumetric data. Along the viewing rays the volume is resampled at evenly located sample points and the samples take the value of the nearest voxel. When the first sample with a value of one is found the corresponding pixel color is determined by shading the intersected surface point.

Then in 1988, Levoy \cite{Levoy1988} published a paper with a raycasting algorithm which since has become the definition for raycasting. Direct volume rendering of gray level volumes is not restricted to surface shaded display like in the case of binary data sets. Here a composite projection of the volume can be performed by evaluating the light transport equation along the viewing rays. Composition requires two important parameters, the color and an opacity at each sample location. Levoy \cite{Levoy1988}  proposed an image order algorithm which assigns these parameters to each grid location in a preprocessing step. The opacity and color values at an arbitrary sample point are calculated by first order interpolation.

In this report, ray casting developed by Levoy  \cite{Levoy1988} is considered as a typical image order algorithm and will be explained in the following section. Images generated by ray casting represent the reference results in terms of image quality, which is shown as Figure~\ref{fig:ImageOrder} :

\begin{figure}[h]  
\centering
    \includegraphics[width=0.4\textwidth]{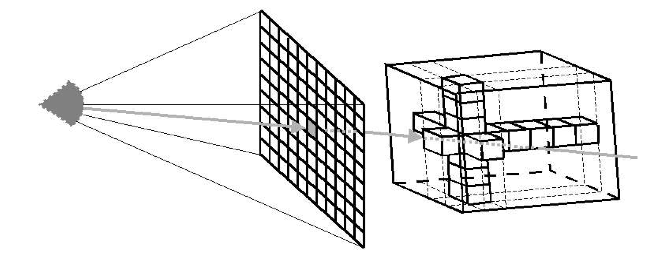}
    \caption{Image Order Rendering}
    \label{fig:ImageOrder}
\end{figure}

\begin{figure}[h]  
\centering
    \includegraphics[width=0.4\textwidth]{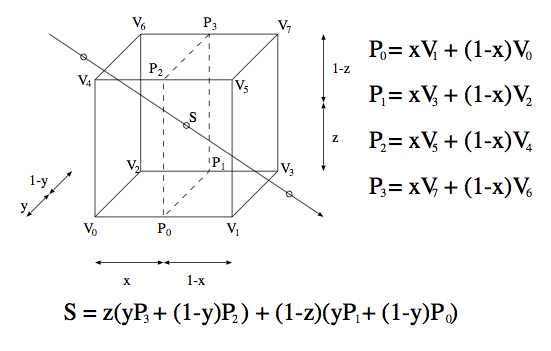}
    \caption{Resampling using trilinear interpolation.}
    \label{fig:alphablending}
\end{figure}

In ray casting, rays are cast into the dataset. Each ray originates from the viewing point, and penetrates a pixel in the image screen, and passes through the dataset. At evenly spaced intervals along the ray, sample values are computed using interpolation [ref: Figure~\ref{fig:alphablending}] . The sample values are mapped to display properties such as opacity and color. A local gradient is combined with a local illumination model at each sample point to provide a realistic shading of the object. Final pixel values are found by compositing the color and opacity values along the ray. The composition models the physical reflection and absorption of light \cite{Ray1999}.  Composite ray casting is a flexible approach for visualizing several semi transparent surfaces contained in the data and produces high quality images. However, the alpha blending evaluation of viewing rays is computationally expensive, especially when super sampling is performed trilinearly interpolating each single sample.

Pseudocode:  Standard Recursive Algorithm \cite{Levoy1988}
\begin{lstlisting}[language={[ANSI]C}, basicstyle=\ttfamily\tiny, numbers=left, numberstyle=\tiny, keywordstyle=\color{blue!70}, commentstyle=\color{red!50!green!50!blue!50}, frame=shadowbox, showstringspaces=false, rulesepcolor=\color{red!20!green!20!blue!20}, xleftmargin=2em,xrightmargin=0em, aboveskip=1em]
For each pixel in image {
Create ray from eyepoint passing through this pixel
 Initialize NearestT to INFINITY and NearestObject to NULL
For every object in scene {
If ray intersects this object {
If t of intersection is less than NearestT { 
Set NearestT to t of the intersection 
Set NearestObject to this object
    } 
  }
}
If NearestObject is NULL {
Fill this pixel with background color
} 
Else {
Shoot a ray to each light source to check if in shadow
If surface is reflective, generate reflection ray: recurse 
If surface is transparent, generate refraction ray: recurse 
Use NearestObject and NearestT to compute shading function 
Fill this pixel with color result of shading function
   } 
}
\end{lstlisting}

Another alternative is discrete ray casting or 3D raster ray tracing \cite{Yagel1992} , where the continuous rays are approximated by discrete 3D lines generated by Discrete Bresenham algorithm \cite{Szirmay-Kalos1995} or continuous scan conversion algorithm.  While traditional ray tracers are capable of rendering only objects represented by geometric surfaces, discrete ray casting is also attractive for ray tracing 3D sampled datasets like 3D MRI imaging,  and computed datasets like fluid dynamics simulations, as well as hybrid models in which such datasets are intermixed with geometric models, such as  scalpel superimposed on a CT image, radiation beams superimposed on a scanned tumor, or a plane fuselage superimposed on a computed air pressure \cite{Kaufman1990} . Unlike nonrecursive ray casting techniques, discrete ray casting, which recursively considers both primary and secondary rays can model shadows and reflections for photorealistic imaging. Discrete ray casting offers the use of ray tracing for improved visualization of sampled and computed volumetric datasets  \cite{Yagel1991} .

Pseudocode:  Discrete Ray Casting   \cite{Yagel1991}
\begin{lstlisting}[language={[ANSI]C}, basicstyle=\ttfamily\tiny, numbers=left, numberstyle=\tiny, keywordstyle=\color{blue!70}, commentstyle=\color{red!50!green!50!blue!50}, frame=shadowbox, showstringspaces=false, rulesepcolor=\color{red!20!green!20!blue!20}, xleftmargin=2em,xrightmargin=0em, aboveskip=1em]
algorithm to determine closest object intersected by a ray
      for each object in the scene
        if the ray intersects with the object then
           if the object is closest to the ray source then
             the object is the closest intersected by the ray
           endif
        endif
      endfor
    endalgorithm
\end{lstlisting}

The closest intersection points are stored for each pixel and afterwards an image space depth gradient shading  \cite{Cohen1990}   \cite{Tam1988}  can be performed. Better results can be achieved applying object space shading techniques like normal based contextual shading \cite{Chen1985} \cite{Herman1981}. Normal computation methods based on surface approximation try to fit a linear \cite{Bryant1989} or a biquadratic \cite{Webber1990} \cite{Webber1991}, where function to the set of points that belong to the same iso surface. These techniques take a larger voxel neighborhood into account to estimate the surface inclination.

There are some advantages and disadvantages of raycasting algorithm.  

Advantages of Raycasting: 

\begin{enumerate}
 \item Realistic simulation of lighting, better than scanline rendering or ray casting;\\
 \item Effects such as reflections and shadows, which are difficult to simulate using other algorithms, are a natural result of the ray tracing algorithm; \\
 \item Relatively simple to implement yet yielding impressive visual results.
\end{enumerate}

Disadvantages of Raycasting: 

\begin{enumerate}
 \item Performance is very poor;\\
 \item Scanline algorithms and other algorithms use data coherence to share computations between pixels, while ray tracing normally starts the process anew, treating each eye ray separately; \\
 \item However, this separation offers other advantages, such as the ability to shoot more rays as needed to perform anti-aliasing and improve image quality where needed.Although it does handle inter-reflection and optical effects such as refraction accurately; \\
\item Other methods, including photon mapping, are based upon ray tracing for certain parts of the algorithm, yet give far better results.
\end{enumerate}

\subsubsection{Object order methods}
Object order algorithms start with a single voxel and compute its contribution to the final image. This task is iteratively performed for all voxels of the data set. Object order rendering is also called forward rendering, or object space rendering or voxel space projection. It loops through the data samples, projecting each sample onto the image plane,  which is shown as Figure~\ref{fig:ObjectOrder} :

\begin{figure}[h]  
\centering
    \includegraphics[width=0.4\textwidth]{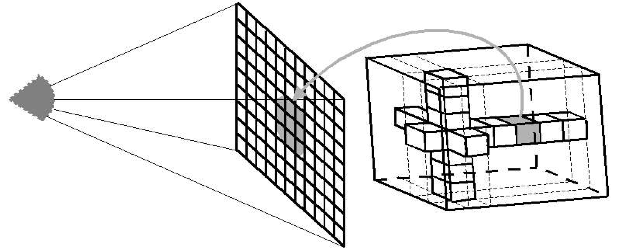}
    \caption{Object Order Rendering}
    \label{fig:ObjectOrder}
\end{figure}

The simplest way to implement viewing is to traverse all the volume regarding each voxel as a 3D point that is transformed by the viewing matrix and then projected onto a Z-buffer and drawn onto the screen. The data samples are considered with a uniform spacing in all three directions. If an image is produced by projecting all occupied voxels to the image plane in an arbitrary order, a correct image is not guaranteed. If two voxels project to the same pixel on the image plane, the one that was projected later will prevail, even if it is farther from the image plane than the earlier projected voxel. This problem can be solved by traversing the data samples in a back to front or front to back order. This visibility ordering is used for the detailed classification of object order rendering.

 The first object order algorithm reported in literature was a rendering method presented by Upson and Keeler \cite{Upson1988} , which processed all voxels in front to back order and accumulated the values for the pixels iteratively.  Similarly to the early image order methods, aimed at the rendering of binary volumes. Processing each sample, only the voxels with value of one are projected onto the screen. The samples are projected in back to front order to ensure correct visibility. If two voxels are projected onto the same pixel, the first processed voxel must be farther away from the image plane. This can be accomplished by traversing the data plane by plane, and row by row inside each plane. For arbitrary orientations of the data in relation to the image plane, some axes may be traversed in an increasing order, while others may be considered in a decreasing order. The ordered traversal can be implemented with three nested loops indexing $x-$,$y -$, and $z-$ directions respectively. Such an implementation supports axis-parallel clipping planes. In this case, the traversal can be limited to a smaller rectangular region by simply modifying the bounds of the traversal. The depth image of the volume can be easily generated. Whenever a voxel is projected onto a pixel, the pixel value is overwritten by the distance of the given voxel from the image plane. Similarly to the early image-based methods the distance image can be passed to a simple 2D discrete shader.

Further development of this idea lead to splatting \cite{Westover1989}
\cite{Westover1990} \cite{Westover1991} \cite{Mueller1996} are not restricted to the rendering of binary volumes. Splatting is an algorithm which combines efficient volume projection with a sparse data representation. In splatting, each voxel is represented as a radially symmetric interpolation kernel, equivalent to a sphere with a fuzzy boundary. Projecting such a structure generates a so called footprint or splat on the screen. Splatting traditionally classifies and shades the voxels prior to projection.

Many many improvements since Westover's approach was published. Crawfis introduced textured splats, Swan and Mueller solved anti alisasing problem, Mueller himself developed  image aligned sheet based splatting and post classified splatting in 1998 and 1999 respectively  \cite{Mueller1999} . Object order approaches also comprise cell projection \cite{Wilhelms1991} and 3D texture mapping.

In Mueller's approach, a gray scale volume is treated as a 3D discrete density function.  Similarly to the ray casting method a convolution kernel defines how to reconstruct a continuous function from the density samples. In contrast, instead of considering how multiple samples contribute to a sample point, it is considered how a sample can contribute to many other points in space. For each data sample $s=(x_{s}, y_{s}, z_{s})$ , a function $C$  defines its contribution to points noted as $(x, y, z)$ in the space: 
\begin{equation}
C(x,y,z)=h(x-x_{s}, y-y_{s}, z-z_{s})*f(s)
\end{equation}

where $f(s)$ is the density of sample $s$. The contribution of a sample $s$ to an image plane pixel $(x,y)$ can be computed by an integration:
\begin{equation}
C(x,y)=f(s)*\int h(x-x_{s}, y-y_{s}, u)du
\end{equation}

where $u$  coordinate axis is parallel to the viewing ray. Since this integral is independent of the sample density, and depends only on its $(x,y)$ projected location, a footprint function $F$can be defined as follows:
\begin{equation}
F(x,y)=\int h(x-x_{s}, y-y_{s}, u)du
\end{equation}

where $(x,y)$ is the displacement of an image sample from the center of the sample’s image plane projection. The footprint kernel $F$ is a weighting function which defines the contribution of a sample to the affected pixels. A footprint table can be generated by evaluating the integral, on a grid with a resolution much higher than the image plane resolution. All the table values lying outside of the footprint table extent have zero weight and therefore need not be considered when generating an image. A footprint table for data sample $s$ is centered on the projected image plane location of $s$ , and sampled in order to determine the weight of the contribution of to each pixel on the image plane.

Computing a footprint table can be difficult due to the integration required. Although discrete integration methods can be applied to approximate the continuous integral, generating a footprint table is still a costly operation. However, in case of orthographic projection, the foot- print table of each sample is the same except for an image plane offset. Therefore, only one footprint table needs to be calculated per view. Since this would require too much computation time anyway, only one generic footprint table is built for the kernel. For each view, a view transformed footprint table is created from the generic footprint table. The generic footprint table can be precomputed, therefore it does not matter how long the computation takes.

Pseudocode:  Splatting an i-axis row of a Data Slice  \cite{Mueller1999}
\begin{lstlisting}[language={[ANSI]C}, basicstyle=\ttfamily\tiny, numbers=left, numberstyle=\tiny, keywordstyle=\color{blue!70}, commentstyle=\color{red!50!green!50!blue!50}, frame=shadowbox, showstringspaces=false, rulesepcolor=\color{red!20!green!20!blue!20}, xleftmargin=2em,xrightmargin=0em, aboveskip=1em]
/* screen coords of transformed point */
float scr[2];
/* screen space step for each step along i-axis of data */
float step[2];
/*pointer to footprint array*/
float *foot;
float alpha, thresh;
/*first and last on-screen points in row*/
int firstpt, lastpt;
/*index into data set*/
int index;
/*pixel address offsets for kernel loop*/
int off[];
/*data set axes indices*/
int i, j, k;
/*lowest pixel coords of footprint coverage */
int lo[2];
/*size of kernel*/
int ksize;
int lopix, p;
/* screen position of first point */
scr[] = transformed(k, j, firstpt);
/* dda increment for each i-step */
step[] = f(view_tansform);
/* index of first point in j'th row and k'th slice */
index = f(k, j, firstpt); 
/* do all on-screen points in row */
for (i = firstpt; i <= lastpt; i++) {
/* load data point value */
alpha = data[index++];
/* test for significant data */
if (alpha > thresh) {
/* footprint is function of fractional screen position */
foot = f(scr[]);
/* find lowest pixel of footprint coverage */
lo[] = (int)(scr[] - extent);
/* index to lowest pixel */
lopix = f(lo[]);
/* loop over kernel size */
for (p = 0; p < ksize; p++) { 
/* accumulate contribution */
image[lopix + off[p]] += alpha * foot[p]; 
  }
}
/* step to position of next point on screen */
scr[] += step[];
}
\end{lstlisting}

There are three modifiable parameters of the splatting algorithm which can strongly affect the image quality. First, the size of the footprint table can be varied. Small footprint tables produce blocky images, while large footprint tables may smooth out details and require more space. Second, different sampling methods can be used when generating the view-transformed footprint table from the generic footprint table. Using a nearest neighbor approach is fast, but may produce aliasing artifacts. On the other hand, using bilinear interpolation produces smoother images at the expense of longer rendering times. The third parameter which can be modified is the reconstruction filter itself. The choice of, for example, a cone function, Gaussian function, Sinc function or bilinear function affects the final image.

Alpha blending composition is also supported by the splatting algorithm. The voxel con- tributions of slices mostly perpendicular to the viewing direction are evaluated on associated sheets parallel to the image plane. After having each sheet generated image composition is performed applying the Porter-Duff over operator.

Using the splatting algorithm approximately the same image quality can be achieved as applying a composite ray casting. The advantages of splatting over ray casting are the following. First, the cache coherency can be exploited since the voxels are sequentially traversed in the same order as they are stored in memory. In contrast, ray casting requires random access to the voxels. Furthermore, the splatting approach supports incremental rendering in back to front or front to back order. Using splatting smooth surfaces can be rendered without staircase artifacts, unlike in the case of ray casting. The main drawback of splatting is that the generated images are blurred because of the spherically symmetric reconstruction kernel. In contrast, using ray casting with trilinear reconstruction sharp object boundaries are obtained.

Advantages of Splatting: 

\begin{enumerate}
 \item Footprints can be pre-integrated, which ensured fast voxel projection;\\
 \item Fast: voxel interpolation is in 2D on screen; \\
 \item More accurate integration (analytic for X-ray);\\
 \item More accurate reconstruction (afford better kernels);\\
 \item Only relevant voxels must be projected. 
\end{enumerate}

Disadvantages of Splatting: 

\begin{enumerate}
 \item Mathematically, the early splatting methods only work for X-ray type of rendering, where voxel ordering is not important, bad approximation for other types of optical models; \\
 \item Object ordering is important in volume rendering, front objects hide back objects
need to composite splats in proper order, else we get bleeding of background objects into the image (color bleeding); \\
 \item However, this separation offers other advantages, such as the ability to shoot more rays as needed to perform anti-aliasing and improve image quality where needed.Although it does handle inter-reflection and optical effects such as refraction accurately; \\
\item Axis aligned approach add all splats that fall within a volume slice most parallel to the image plane, composite these sheets in front to back order, incorrect accumulating on axis aligned face cause popping; \\
\item A better approximation with Riemann sum is to use the image aligned sheet based approach. 
\end{enumerate}

\section{Acceleration Techniques}

Early implementations of volume rendering used brute-force techniques that require on the order of 100 seconds to render typical data sets on a workstation. Algorithms with optimizations that exploit coherence in the data have reduced rendering times to the range of ten seconds but are still not fast enough for interactive visualization applications.

Many of the three dimensional data sets that need to be visualised contain an interesting range of values throughout the volume. By interesting, it is meant those parts of the volume to which the viewer’s attention must be drawn in order for the viewer to gain insight to the physical phenomena the data represents. If the range of values is small, as for example the visualisation of the human skull from CT scans, then a surface tiling method will suffice.

Volume rendering offers an alternative method for the investigation of three dimensional data, such as  surface tiling as described by Jones \cite{Jones1995} , marching cubes supported iso surface rendering by Lorensen and Cline \cite{Lorensen}, octree acceleration for faster iso surface generation by Wilhelms and Van Gelder \cite{Wilhelms1992}, special data structure for rendering by Wyvill et. al. \cite{Wyvill1986} and surface mapping mathod by Payne and Toga \cite{Payne1990}.

 Surface tiling can be regarded as giving one particular view of the data set, one which just presents all instances of the threshold value.  All other values within the data are ignored and do not contribute to the final image. This is acceptable when the data being visualised contains a surface that is readily understandable, as is the case when viewing objects contained within the data produced by CT scans. In certain circumstances this view alone is not enough to reveal the subtle variations in the data, and for such data sets volume rendering was developed \cite{Levoy1988} \cite{Sabella1988} \cite{Upson1988} \cite{Carpenter1988}.

Most data sets do not fall into this category, but rather have a larger range of values or several different values which need to be represented in the visualisation. Such data sets need a method which can display the volume as a whole and visualise correctly those data values in which the viewer is interested.

There are several widely used optimization methods, for early ray termination and empty space skipping. Early ray termination compares accumulated opacity against threshold, such as marching cubes algorithm \cite{Lorensen}, and in such a way accelerates rendering process.  Empty space skipping method utilize additional data structure \cite{Wyvill1986}, and encoding empty space in volume, such as Octree algorithm \cite{Wilhelms1992}, which encodes measure of empty within 3D texture read from fragment shader, and performs raymarching fragment shader can modulate sampling distance based on empty space value.

In this report, we will present the fast image order and object order methods respectively. It will be shown that the advantageous properties of these two different approaches are complementary. Therefore, hybrid methods which combine image order and object order techniques have been proposed by several authors will be descibed right after.  One of them is a two pass raycasting and back projection algorithm which exploits the frame to frame coherency. Another one is the classical shear warp algorithm, which is based on run length encoding of volume and image scanlines exploiting the volume and image coherency respectively.

\subsection{Fast image order techniques}
\subsubsection{Hierarchical data structures}
Hierarchical data structures like octrees,  k-d trees, or pyramids are used for image order volume rendering to efficiently encode the empty regions in a volume. Such data structures are widely used in computer graphics for accelerating traditional algorithms, like ray tracing. Among which, the most widely used method is octree. The use of octrees for 3D computer graphics was pioneered by Donald Meagher at Rensselaer Polytechnic Institute in 1980. \cite{Meagher1980}   The idea of using octree is to quickly find the first intersection point for an arbitrary ray without evaluating the intersections with all the objects. Ray tracing in continuous analytically defined scenes requires a hierarchical struc- ture with arbitrarily fine resolution. In contrast, in volume rendering the discrete representation of the scene can be exploited.  Then a pointerless complete octree represented by a pyramid was introduced by Levoy \cite{Levoy1990} \cite{Wilhelms1990}.

Assuming that the resolution of the volume is $N*N*N$, where $N=2^{M}+1$ for some integer $M$ . A pyramid is defined as a set of $M+1$ volumes. Volumes are indexed by a level number $m=0,1,2,...,M$ , and the volume at level $m$ is denoted by $V_{M}$ .  Volume $V_{0}$ measures $N-1$ cells on a side, volume $V_{1}$ measures $(N-1)/2$ cells on a side and so on up to the volume $V_{M}$ which is a single cell.

Levoy applied a binary pyramid in order to quickly traverse the empty regions in a volume.
 \cite{Levoy1990}  A cell of $V_{0}$ represents the rectangular regions between eight neighboring voxels in the original data. The value of a cell in $V_{0}$ is zero if all of its eight corner voxels have opacity equal to zero, otherwise its value is one. At higher levels of the pyramid zero is assigned to a cell if all the corner cells at one level lower have value of zero.

For each ray, first the point where the ray enters a single cell at the top level is calculated. Afterwards the pyramid is traversed in the following manner. Whenever a ray enters a cell its value is checked. If it contains zero the ray advances to the next cell at the same level. If the parent of the next cell differs from the parent of the previous one then the parent cell is investigated and the ray is traced further at one level higher. If the parent cell is empty then it can be skipped, and the iteration continues until a non-empty cell is found. In this case, moving down in the hierarchical structure, the first elementary cell is determined which has at least one opaque corner voxel. In such an elementary cell samples are taken at evenly spaced locations along the ray and compositing is performed. Using such a hierarchical ray traversal larger empty regions can be easily skipped. Since the non-empty cells in the binary pyramid represent the regions, where opaque voxels are present the algorithm is called presence acceleration.

In 1992, Denskin and Hanrahan \cite{Danskin1992} improved this algorithm using pyramids not only for skipping the empty ray segments but for approximate evaluation of homogeneous regions. Therefore, their technique is called homogeneity acceleration. Instead of using a binary pyramid they construct a so called range pyramid which contains the maximum and minimum values of subvolumes at one level lower. If the maximum and minimum values of a cell are nearly the same then it is considered homogeneous and an approximate evaluation is performed.

\subsubsection{Early ray termination}

To reduce the time complexity of volume rendering, Leovy \cite{Levoy1988} came up with a technique named as early ray termination (ERT), which adaptively terminates accumulating color and opacity values in order to avoid useless ray casting. This technique reduces the execution time by roughly a factor of between 5 and 11. 

ERT reduces the computational amount by avoiding accumulation of color and opacity values that do not have influence on the final image.  Associating an accumulated opacity to each pixel of the image plane ray casting can be performed evaluating the rays in front to back order, in case of back to front composition, all the samples along the ray have to be taken into account. This computation is usually redundant since several samples can be occluded by a ray segment which is closer to the viewer and has accumulated opacity of one. Therefore, these samples do not contribute to the image. In contrast, using front to back composition, the rays can terminate when the accumulated opacity exceeds a predefined threshold \cite{Levoy1990} . This technique is well known as early ray termination or $\alpha$ acceleration.

This acceleration method introduces a systematic bias in the image because of the predefined
threshold. In order to avoid this, a technique called Russian Roulette can be used for unbiased estimation of pixel values \cite{Danskin1992}.

Earlier parallel schemes can be classified into two groups: screen parallel and object parallel rendering as illustrated in Figure~\ref{fig:Raycastingschemes} :

\begin{figure}[h]  
\centering
    \includegraphics[width=0.4\textwidth]{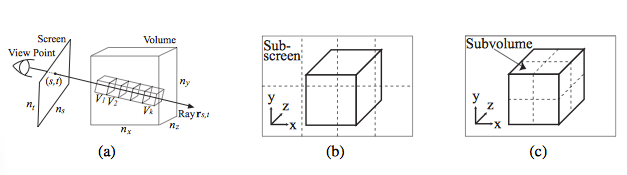}
    \caption{Ray casting and its parallel schemes : (a) ray casting by exploiting the parallelism in screen space and in object space (b) Screen-parallel rendering and (c) object-parallel rendering parallelize, respectively.}
    \label{fig:Raycastingschemes}
\end{figure}

\begin{figure}[h]  
\centering
    \includegraphics[width=0.4\textwidth]{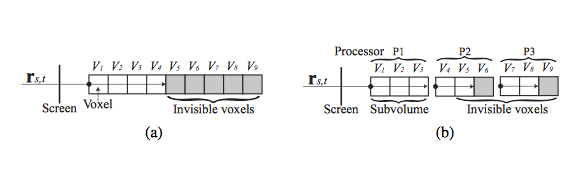}
    \caption{Early ray termination (ERT). (a) Global ERT for sequential and screen-parallel rendering, and (b) local ERT for object-parallel rendering. While global ERT terminates the ray immediately before invisible voxel $V_{5}$, local ERT fails to avoid accumulating locally visible but globally invisible voxels: $V_{5}$, $V_{7}$, and $V_{8}$. Voxels $V_{6}$ and $V_{9}$ are invisible locally as well as globally.
}
    \label{fig:ERT}
\end{figure}

Screen parallel rendering exploits the parallelism in screen space. In this scheme, the screen is divided into p subscreens, where p represents the number of processors, and tasks associated with each subscreen are assigned to processors. Because each processor takes responsibility for the entire of a ray as it does in sequential schemes, ERT can easily be applied to this scheme, as illustrated in Figure~\ref{fig:ERT} (a). Furthermore, by assigning the tasks in a cyclic manner, this scheme statically balances the processing workloads. However, it requires large main memory to provide fast rendering for any given viewpoint, because every processor need to load the entire volume into memory. Thus, though screen parallel rendering is a good scheme for small datasets, which require no data decomposition, it does not suit for large scale datasets.

In contrast, object parallel rendering exploits the parallelism in object space. This scheme divides the volume into p subvolumes, and then assigns tasks associated with each subvolume to processors. Parallel rendering of each subvolume generates p dis- tributed subimages, so that image compositing is required to merge subimages into the final image. Thus, this scheme allows us to distribute subvolumes to processors, so that is suitable for large scale datasets. However, because accumulation tasks of a ray can be assigned to more than one processor, it is not easy to utilize global ERT in this scheme.

Figure~\ref{fig:ERT} shows an example of local ERT in object-parallel rendering. In this example, voxels from $V_{1}$ to $V_{4}$ are visible from the viewpoint while voxels from $V_{5}$ to $V_{9}$ are invisible. These voxels are assigned to three processors, so that each processor takes responsibility for three of the nine voxels. In object parallel rendering, the reduction given by ERT is localized in each processor, because processors take account of the local visibility instead of the global visibility. 

\subsubsection{Distance transformation}

The main drawback of acceleration techniques based on hierarchical data structures is the additional computational cost required for the traversal of cells located at different levels of the hierarchy. Furthermore, the rectangular cells only roughly approximate the empty regions.

Cohen proposed a technique called proximity clouds for fast 3D grid traversal. Here the geometric information required for empty space skipping is available with the same indices used for the original volume data \cite{Cohen1994}. The data structure is a simple 3D distance map generated from binary volumes. The input binary volume encodes the transparent and opaque cells similarly to Levoy’s approach \cite{Levoy1990}. In the distance volume each cell contains the distance to the nearest non-empty cell. The distance between two cells is represented by the Euclidean distance of their center points.

Ray casting is performed applying two different cell traversal strategies. The algorithm switches between these two strategies depending on the distance information stored in the current cell. Using fixed-point arithmetic and integer division it is easy to find the cell which contains the current ray sample. If the current sample is in the vicinity of an object a simple incremental ray traversal is performed.

If this is not the case, the distance value $d$ stored in the current cell is used for fast skipping of empty regions. The new sample is determined by adding the unit direction of the given ray multiplied by $d-1$ to the current sample location. The distance from the nearest object has been calculated from the center of the current cell, therefore using a stepping distance $d-1$ , skipping beyond the free zone can be avoided.

Distance maps can be generated based on several distance metrics, like City Block, Euclidean, or Chessboard distance. Approximate distance maps are usually calculated applying the efficient Chamfering method. The basic idea is to use a mask of local distances and propagate these distances over the volume. The generation of a more accurate distance map requires a larger mask, therefore the preprocessing time is longer. On the other hand, less samples have to be taken in the raycasting process when more exact distance information is available. Therefore, the applied distance metric is a compromise between the preprocessing and rendering times.

\subsection{Fast object-order techniques}

\subsubsection{Hierarchical splatting}

Object-order volume rendering typically loops through the data, calculating the contribution of each volume sample to pixels on the image plane. This is a costly operation for high resolution data sets. One possibility is to apply progressive refinement. For the purpose of interaction, first a lower quality image is rendered. This initial image is progressively refined when a fixed viewing direction has been selected.

For binary data sets, bits can be packed into bytes such that each byte represents a $2*2*2$
portion of the data \cite{Tuy1984} . The volume is processed bit by bit to generate the full resolution image but lower resolution images can be produced processing the volume byte by byte. A byte is considered to represent an element of an object if it contains more than four non zero bits, otherwise it represents the background. Using this technique, an image with one half the linear resolution is produced in approximately one eight the time.

A more general method for decreasing data resolution is to build a pyramid data structure, which for on original data set of $N^{3}$ samples, consists of a sequence of $log(N)$ olumes. The first volume is the original data set, while the second volume of one-eighth the resolution is created by averaging each $2*2*2$ group of samples of the original data set. The higher levels of the volume pyramid are created from the lower levels in a similar way until  $log(N)$ olumes have been created. An efficient implementation of the splatting algorithm, called hierarchical splatting \cite{Laur1991} uses such a pyramid data structure. According to the desired image quality, this algorithm scans the appropriate level of the pyramid in a back to front order. Each element is splatted onto the image plane using the appropriate sized splat. The splats themselves are approximated by polygons which can be rendered by conventional graphics hardware.

Laur and Hanrahan \cite{Laur1991} introduced hierarchical splatting for volume rendering using Gouraud shaded polygons. Researchers like Mueller \cite{Mueller1999} , Swan \cite{Swan1997}, and Zwicker \cite{Zwicker2001} focus mainly on the improvement of the visual quality of texture splatting; however, the techniques described in these papers only apply to the reconstruction of continuous functions, take  volume rendering of regular grid data for example, and they do not address adaptive ren- dering or data size reduction. Additionally, there exist a number of non-realtime rendering systems for large point based data sets, e.g. for rendering film sequences \cite{Cox1996} .

Using points as rendering primitives is a topic of ongoing research. However, almost all publications in this area deal with the rendering of geometric surfaces. Alexa \cite{Alexa2001} , Pfister \cite{Pfister2000} , Rusinkiewicz and Levoy \cite{Rusinkiewicz2000} , Wand \cite{Wand2001} , and Zwicker \cite{Zwicker2001} showed different methods to create data hierarchies of surfaces represented by sample points and how to render them efficiently. As the intrinsic model of points describing a surface is fundamentally different to the model used for scattered data, their clustering techniques cannot be applied in our case. Pauly \cite{Pauly2002} used principal component analysis for clustering, but with a different hierarchy concept compared to our approach. Some systems \cite{Rusinkiewicz2000}  \cite{Botsch2002} use quantized relative coordinates for storing the points in a hierarchical data structure, but these approaches were not optimized for fast GPU access because the data structures had to be interpreted by the CPU. Additionally, the presented rendering techniques have been designed to create smooth surfaces without holes and they allow no or only few layers of transparency.

\subsubsection{Extraction of surface points}

Although there are several optimization techniques based on empty space leaping or approximate evaluation of homogeneous regions, because of the computationally expensive alpha blending compositing the rendering is still time demanding.

One alternative to alpha blending volume visualization is the extraction of relevant voxels and the optimization of the rendering process for sparse data sets. Following these approach interactive applications can be developed which support flexible manipulation of the extracted voxels. In medical imaging systems, for example, the cutting operations are rather important, where a shaded isosurface and an arbitrary cross sectional slice can be rendered at the same time.

Sobierajski \cite{Sobierajski1993} proposed a fast display method for direct rendering of boundary surfaces. From the volume data only those boundary voxels are extracted which are visible from a certain set of viewing directions. In many cases the six orthographic views are sufficient to obtain an approximate set of all the potentially visible boundary voxels. Better approximation can be achieved increasing the number of directions.

Taking only the six orthographic views into account the visibility calculations can be performed efficiently using a 2D boundary tracing algorithm on the slices perpendicular to each coordinate axis. The output of the surface extraction algorithm is a list of boundary voxels in which duplicate elements are removed. The generated list stores for each voxel all the attributes which are necessary for the rendering process, like the coordinates or the normal vector.

The set of surface points is passed to the rendering engine. Since adjacent voxels are mapped onto not necessarily adjacent pixels, holes can appear in the produced image. In order to avoid this problem one voxel is projected onto several pixels depending upon the viewing direction. Since the rendering speed is directly related to the length of the voxel list, for a specific viewing direction the number of voxels to be projected can be reduced by voxel culling. This is similar to back face culling in polygon rendering \cite{Szirmay-Kalos1995} . If the dot product of the surface normal and the viewing direction is positive the given voxel belongs to a back face, therefore it is not rendered.

Since the presented algorithm follows a direct volume rendering approach, cutting planes can be easily implemented. The projected boundary surface points are shaded according to the lighting conditions and the voxels intersected by the cutting plane are rendered by projecting their original density values onto the image plane.

A fast previewing algorithm proposed by Saito \cite{Saito1994} is also based on the extraction of boundary voxels. Similarly to the previous method a set of surface points is stored in a list and it is passed to the rendering engine. In contrast, in order to increase the rendering speed, only a subset of the boundary voxels are extracted according to a uniform distribution. The extracted surface samples are converted to geometrical primitives like crosslines perpendicular to the surface normal and projected onto the image plane.

\subsection{Hybrid acceleration methods}

\subsubsection{Shear-warp factorization}
The shear-warp algorithm is a purely software based renderer. Shear-warp was invented by Lacroute  \cite{Lacroute1995} and can be considered a hybrid between image order algorithms, such as raycasting, and object-order algorithms, such as splatting. In shear warp, the volume is rendered by a simultaneous traversal of run length encoding (RLE) encoded voxel and pixel runs, where opaque pixels and transparent voxels are efficiently skipped during these traversals. Further speed comes from the fact that a set of interpolation weights is precomputed per volume slice and stays constant for all voxels in that slice. The caveat is that the image must first be rendered from a sheared volume onto a so called base plane, aligned with the volume slice most parallel to the true image plane Figure~\ref{fig:shearwarp} :

\begin{figure}[h]  
\centering
    \includegraphics[width=0.4\textwidth]{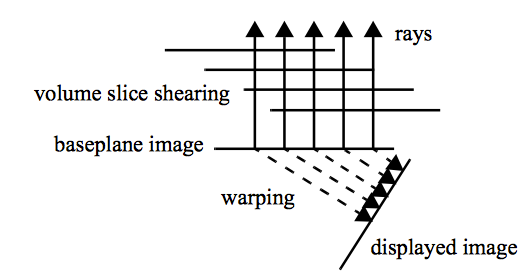}
    \caption{A sketch of the shear-warp mechanism}
    \label{fig:shearwarp}
\end{figure}

 After completing the base plane rendering, the base plane image is warped onto the true image plane and the resulting image is displayed.

Pseudocode:  Pseudocode of the standard shear warp algorithm.
\begin{lstlisting}[language={[ANSI]C++}, basicstyle=\ttfamily\tiny, numbers=left, numberstyle=\tiny, keywordstyle=\color{blue!70}, commentstyle=\color{red!50!green!50!blue!50}, frame=shadowbox, showstringspaces=false, rulesepcolor=\color{red!20!green!20!blue!20}, xleftmargin=2em,xrightmargin=0em, aboveskip=1em]
Shear_Warp (voxel_data)
 if New(Mview)
shade_table = Construct_Shade_Cube(Mview, L, E); 
if New(transfer_function)
RLE_Encode(voxel_data, RLE_X, RLE_Y, RLE_Z);
if (major_viewing_axis == X) current_RLE = RLE_X;
if (major_viewing_axis == Y) current_RLE = RLE_Y;
if (major_viewing_axis == Z) current_RLE = RLE_Z; 
Factorize(Mview, Mshear, Mwarp);
ShearParameters(Mshear, shear_u, shear_v, trans_u, trans_v); 
base_image = Render(current_RLE);
display_image = Warp(base_image, Mwarp);


Render (current_RLE)
 base_image.Initialize(); 
num_scanline_pixels = slice_width; 
num_scanlines = slice_ height;
for k = front_slice, k<=end_slice, k++ 
Composite_Slice (k);
return(base_image);


Composite_Slice (k)
slice_u =k.shear_u + translate_u;
slice_v = k.shear_v + translate_v;
slice_u_int=floor(slice_u);
slice_v_int =floor(slice_v);
weights[4]=Voxel_Weights(u, u_int, v, v_int); 
for j= 0, j<=num_scanlines-1, j++
for i= 0, i<= num_scanline_pixels-1, i++
bot_run=Skip_Transparent_Voxels (i, j, k, current_RLE); 
top_run=Skip_Transparent_Voxels (i, j+1, k, current_RLE);
pixel_start=Min (bot_run.start_voxel, top_run.start_voxel);
a=bot_run.start_voxel + bot_run.length;
b=top_run.start_voxel + top_run.length;
pixel_end=Max (a,b); 
for pixel=pixel_start, pixel<=pixel_end, pixel++
pixel =Skip_Opaque_Pixels (pixel, j, base_plane_image.opacity_RLE)
 if (pixel > pixel_end)
break;
voxel_square=Get_Voxel_Square (bot_run, top_run, pixel); 
composited_opacity=Composite_Pixel (voxel_square, weights);
if composited_opacity > 0
Update_Opacity_RLE (pixel, base_plane_image.opacity_RLE); 
i=pixel;


RLE_Encode (voxel_data, RLE_X, RLE_Y, RLE_Z)
for all voxels in voxel_data with AlphaTF[voxel.density] > 0
RLE_voxel.shade_index=Calc_Shade_Index (voxel.gradient); 
RLE_voxel.density=voxel.density;
RLE_X.Add (RLE_X, RLE_voxel);
RLE_Y.Add (RLE_voxel);
RLE_Z.Add(RLE_voxel);


Composite_Pixel (voxel_square, weights)
voxel_square.opacities=AlphaTF[voxel_square.densities]; 
pixel_opacity=Interpolate (voxel_square.opacities, weights);
if pixel_opacity > 0
voxel_square.shades=
   Get_Shades (shade_table, voxel_square.shade_indices); 
voxel_square.colors=
   Calc_Colors (voxel_square.shades, voxel_square.densities, TF);
pixel_color=Interpolate (voxel_square.colors, weights);
composited_opacity=
   Composite (color, opacity, base_plane_image.pixels);
return (composited_opacity);

\end{lstlisting}

\section{Challenges of Medical Volume Rendering}

In the past decade, commercial CT scanners have become available that can take five 320 slice volumes in a single second.  Toshiba's 320 slices CT scanner, the Aquilion One, was introduced in 2007. \cite{Hsiao2010} This is fast enough to make 3D videos of a beating heart.  Rapid advances in the dynamic nature and sheer magnitude of data force us to make improvements to existing techniques of medical visulization to increase computational and perceptual scalability.

Images of  erve bundles and muscle fibres is improtant for areas of study in neuroscience and biomechanics. High Angular Resolution Diffusion Imaging (HARDI) \cite{Tuch2002} and Diffusion Spectrum Imaging (DSI) \cite{Hagmann2006} datasets contain hundreds of diffusion-weighted volumes describing the diffusion of water molecules and hence indirectly the orientation of directed structures, which are calling for new visualization techiniques. 

Then there are the imaging techniques that work on the level of molecules and genes.  Up to now, most of the visualization research has been focused on small animal imaging \cite{Kok2010} \cite{Kok2007} , but due to its great diagnostic potential, molecular imaging will see increased application in humans. The great potential of these is that they can reveal pathological processes at work long before they become apparent on the larger scale, such as bioluminescence (BLI) and fluorescence (FLI) imaging's applications on tumor detection.

Acquiring the image data is just one part of the challenge. Representing it visually in a way that allows the most effective analysis is also hugely difficult but again there have been huge advances.

One of the most spectacular is the representation of medical data topologically, in other words showing the surfaces of objects. That makes it possible to more easily see the  shapes of organs and to plan interventions such as surgery.

Another of the most spectacular is the interactive representation of multi-modality medical data. Medical visualization research often combines elements of image analysis, graphics and interaction, and is thus ideally equipped to address the challenge of developing and validating effective interactive segmentation approaches for widespread use in medical research and practice. Also, integration of simulation models and MRI or CT is crucial for diagnostics as well.  A huge challenge for the future and the subject of much current research, is to create images of the potential outcome of interventions that show the result of the surgery.

The most recent image processing techniques allow the addition of realistic lighting effects creating photo realistic images. Beyond this, hyper realistic images can show what lies beneath certain layers.

Another area of growing importance is the visualisation of multi-subject data sets, this is required as researchers want to study the onset and progression of disease, general aging effects, and so forth in larger groups of people. such as the Rotterdam Scan Study focusing on neuro degeneration \cite{Leeuw2001} and the Study of Health In Pomerania (SHIP) focusing on general health \cite{John2001},  Steenwijk setting the first steps for the visualization of population imaging by applying visual analysis techniques to cohort study imaging data \cite{Steenwijk2010}.

Last but not least, with increasing population of mobile device, such as iPhone and iPad, cheaper and easy compiled visualization softwares are needed for medical purposes, to free doctors from desktops. The biggest challenge of all is to find ways of making powerful medical visualisation techniques cheap enough for everyone.

\phantomsection

\section*{Acknowledgments} 

\addcontentsline{toc}{section}{Acknowledgments} 

Thanks for Gelman library  and Prof. Hahn for supporting. 

\phantomsection


\end{document}